\documentclass[aps,preprint,amsmath,amssymb]{revtex4}
\usepackage{graphicx,axodraw}

\newcommand{\bm}[1]{\mbox{\boldmath{$\rm #1$}}}

\begin{document}

\title
{\Large \bf Leptogenesis in the extension of the
Zee-Babu~model}

\author{
Chian-Shu~Chen\footnote{e-mail: chianshu@gmail.com},
Chao-Qiang~Geng\footnote{e-mail: geng@phys.nthu.edu.tw} and
Dmitry~V.~Zhuridov\footnote{e-mail: zhuridov@phys.nthu.edu.tw}}
\affiliation{Department of Physics, National Tsing Hua University,
Hsinchu, Taiwan 300}

\date{\today}

\begin{abstract}
We demonstrate  that the  extension of the Zee-Babu model
can generate not only the small neutrino masses but also the baryon number asymmetry
in the universe. In particular, we show that the scale of the singlet scalar responsible for the leptogenesis
can be of order 1~TeV, that can be tested at the LHC and ILC. We also considered the possible minimal extension of this model to generate the dark matter.
\end{abstract}

\maketitle

The zero neutrino masses predicted in the standard model (SM) turn
out to be inconsistent with the experimental data~\cite{PDG}. The
problem can be solved by including right-handed neutrinos $\nu_R$
to the particle content~\cite{seesaw}. On the other hand, without
$\nu_R$, many theories have been proposed to explain the small
active neutrino masses by using extended scalar
fields~\cite{ExtendedHiggs,Zee-Babu,CGN,CGZ}. For example, in the
Zee-Babu (ZB) model~\cite{Zee-Babu}, it contains only one singly
and one doubly charged $SU(2)$ singlet scalars beside the SM Higgs
doublet. The Majorana neutrino masses arise radiatively at
two-loop level. However, it is clear that the original ZB model
cannot accomplish the leptogenesis~\cite{LG,BPY}.

In this paper, we would like to  extend the ZB model slightly by
including two additional doubly charged and one neutral singlet scalars in order to
generate a possible lepton number asymmetry in the early
universe, which can be converted by the usual sphaleron mechanism
into the present baryon number asymmetry~\cite{BA,Harvey_Turner}.
In this model we will demonstrate that the $CP$
asymmetry can be induced by the interference of tree and one-loop
contributions in the three body decays of the neutral scalars to dileptons and
doubly charged Higgs (DCH). Moreover, we show that the
experimental constraints on the
parameters~\cite{Nebot} leave a window for the TeV scale
leptogenesis that gives the opportunity to test the model at the Large Hadron Collider (LHC) and International Linear Collider (ILC)~\cite{CGN,CGZ,ILC,ILC1}. The degeneracy of the DCH masses is not required. This is an attractive feature of three-body decay mechanisms of the leptogenesis~\cite{Hambye}. Note that in the absence of the singlet scalar, the asymmetry generated in the decays of the DCHs 
is washed out by the gauge scattering processes~\cite{Hambye,HMS}. While the presence of neutral singlet scalars can improve a radiative stability of Higgs potential~\cite{ST}.

\vspace{0.8cm}

The scalar sector content and quantum numbers in our extension of the ZB model are listed in Table~\ref{T1},
where $\phi$, $s$, $h$ and $k$ represent the Higgs
doublet, neutral singlet, singly charged singlet and doubly charged singlet
scalars, respectively.
\begin{table}[htb]
\begin{center}
\caption{ The scalar fields, its electro-weak charges and $Z_2$ parity; $i=1,2$.}\label{T1}
\begin{tabular}{|c|c|c|}
  \hline
   Scalar & $SU(2)_L \times U(1)_Y$ & $Z_2$ \\
  \hline
  $\phi$& (2, 1) & + \\
  $s$     & (1, 0) & - \\
  $h$     & (1, 1) & + \\
  $k_i$& (1, 2) & + \\
  $k_3$& (1, 2) & - \\
  \hline
\end{tabular}
\end{center}
\end{table}
The most general scalar potential is written as
\begin{eqnarray}\label{V}
V &=& -\mu_\phi^2|\phi|^2 + \lambda_{\phi}|\phi|^4 + M_s^2s^2 + \lambda_ss^4 + M_h^{02}|h|^2 +
\lambda_h|h|^4 \nonumber\\
&& + M_{k_i}^2|k_i|^2 +
\lambda_{k_i}|k_i|^4 + \lambda_{kij}|k_i|^2|k_j|^2 \nonumber \\
&& + \lambda_{\phi s}|\phi|^2s^2 + \lambda_{\phi h}|\phi|^2|h|^2
+ \lambda_{\phi k_i}|\phi|^2|k_i|^2 + \lambda_{sh}s^2|h|^2 + \lambda_{sk_i}s^2|k_i|^2
 + \lambda_{hk_i}|h|^2|k_i|^2 \nonumber \\
&& + \left[\lambda_{\phi12}|\phi|^2k_1^{\dag}k_2 +
\lambda_{h12}|h|^2k_1^{\dag}k_2 + \mu_{sn3}sk_n^\dag k_3 + \mu_{hn}(h^+)^2k_n^{--} + {\rm H.c.} \right],
\end{eqnarray}
where we assume summation over the repeated indexes, $i,j=1,\cdots,3$ ($i\neq j$) and $n=1,2$.

After the $SU(2)_L\times U(1)_Y$ symmetry
breaking by the SM Higgs VEV of $v$, the doubly charged Higgs mass
matrix is given by
\begin{eqnarray}
{\cal M}^2 = \left(\begin{array}{cc}M_{k_1}^2 +
\frac{\lambda_{\phi k_1}}{2}v^2 & \frac{\lambda_{\phi12}}{2}v^2
\\\frac{\lambda_{\phi12}^*}{2}v^2 & M_{k_2}^2 +
\frac{\lambda_{\phi k_2}}{2}v^2\end{array}\right),
\end{eqnarray}
which can be diagonalized by the orthogonal transformation
\begin{eqnarray}
    \Theta = \left(\begin{array}{cc}\cos{\theta} & \sin{\theta} \\
-\sin{\theta} & \cos{\theta}\end{array}\right)
\end{eqnarray}
to form the mass  eigenstates $P = \Theta k$ with the masses
\begin{eqnarray}
    M_{1,\,2} = (Q\pm\sqrt{Q^2-R})/2,
\end{eqnarray}
where
\begin{eqnarray}
    Q=\sum_{i=1,2} (M_{k_i}^2 + \lambda_{\phi k_i}v^2/2), \quad R
    = 4(M_{k_1}^2 + \lambda_{\phi k_1}v^2/2) (M_{k_2}^2 + \lambda_{\phi
    k_2}v^2/2) - |\lambda_{\phi12}|^2v^4
\end{eqnarray}
with the requirement of $Q^2\geq R\geq 0$.

The third doubly charged $P_3$, singly charged $h$ and neutral $S$ scalar mass eigenstates have squared masses
$M_3^2=M_{k_3}^2 + \lambda_{\phi k_3}v^2$, $M_h^2=M_{h}^{02} + \lambda_{\phi h}v^2$ and $M_S^2=M_s^2 + \lambda_{\phi s}v^2$, correspondingly.

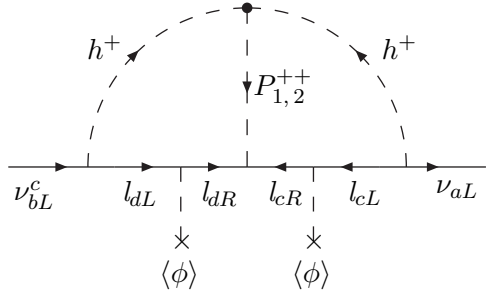
\begin{figure}[ht]
\begin{picture}(300,100)(0,0)
\ArrowLine(80,40)(120,40) \Text(90,35)[t]{$\nu_{bL}^c$}
\ArrowLine(120,40)(145,40) \Text(130,35)[t]{$\emph{l}_{dL}$}
\DashLine(145,40)(145,12){5}\Text(146,12)[c]{$\times$}
\Text(145,5)[t]{$\langle\phi\rangle$} \ArrowLine(145,40)(170,40)
\Text(160,35)[t]{$\emph{l}_{dR}$} \ArrowLine(220,40)(195,40)
\Text(215,35)[t]{$\emph{l}_{cL}$}
\DashLine(195,40)(195,12){5}\Text(196,12)[c]{$\times$}
\Text(195,5)[t]{$\langle\phi\rangle$} \ArrowLine(195,40)(170,40)
\Text(185,35)[t]{$\emph{l}_{cR}$} \ArrowLine(220,40)(260,40)
\Text(250,35)[t]{$\nu_{aL}$} \DashArrowArcn(170,40)(60,180,90){5}
\Text(110,90)[lt]{$h^+$} \DashArrowArc(170,40)(60,0,90){5}
\Text(235,90)[rt]{$h^+$} \DashArrowLine(170,100)(170,40){7}
\Text(173,70)[l]{$P_{1,\,2}^{++}$} \Vertex(170,100){2}
\end{picture}
\caption{Majorana masses of neutrinos at the two-loop
level.}\label{Zee1}
\end{figure}

The two-loop Majorana neutrino mass generation, shown in
Fig.~\ref{Zee1}, takes place due to the simultaneous presence of
the four couplings: the SM Yukawa $\bar LY\tilde\phi\ell_R$, the
non-SM scalar $\mu_i h^2P_i$ ($\mu_i = \Theta\mu_{hk_i}$) in Eq.~(\ref{V}),
and the two non-SM scalar--lepton
\begin{eqnarray}\label{LY}
{\cal L}_Y = f_{ab}\bar{\tilde{L_a}}L_bh^{+} +
h_{iab}\bar\ell_{aR}^c\ell_{bR}P_{i}^{++} + {\rm
H.c.},
\end{eqnarray}
where  $L$ is the $SU(2)_L$ lepton
doublet with $\tilde{L} = i\tau_2L^c = i\tau_2C\bar{L}^T$,
$\ell_R$ denotes the right-handed charged lepton singlet, and
$a,\,b = e,\,\mu,\,\tau$. Without
loss of generality, the Yukawa matrix $Y$ can be chosen to be diagonal with real
and positive elements~\cite{Nebot,Santamaria}. The matrix $f_{ab}$
is antisymmetric due to the Fermi statistics of the lepton doublets, while
$h_{iab}$ is symmetric under
the indexes of $a$ and $b$.

The neutrino mass matrix, defined as an effective term in the
Lagrangian ${\mathcal L}_\nu \equiv -\frac{1}{2}\bar\nu_{aL}M_{\nu
ab}\nu_{bL}^c + {\rm H.c.}$, is given by \cite{CGN,B_M,2loop}
\begin{eqnarray}
\label{Mnu}
M_{\nu ab} &=& 16\sum_{c,\,d}f_{ac}m_c
\left(\mu_1^*h^*_{1cd}I_{1cd} + \mu_2^*h^*_{2cd}
I_{2cd}\right)m_df_{db},
\end{eqnarray}
where $m_a = Y_{aa}v/\sqrt{2}$ is the charged lepton mass,
$a,\,b,\,c,\,d = e,\,\mu,\,\tau$, and
\begin{eqnarray}\label{Icd}
I_{icd} =
\int\frac{d^4k}{(2\pi)^4}\int\frac{d^4q}{(2\pi)^4}\frac{1} {k^2 -
m_c^2}\frac{1}{k^2 - M_h^2}\frac{1}{q^2 - m_d^2}\frac{1} {q^2 -
M_h^2}\frac{1}{(k-q)^2 - M_{i}^2}.
\end{eqnarray}
We remark that  $M_{\nu ab}$ in Eq. (\ref{Mnu}) has enough freedom
to fit the normal or inverted neutrino mass hierarchy like the
original ZB model~\cite{B_M,Giunti}. 

To study the leptogenesis,
the relevant terms in the Lagrangian are given by
\begin{eqnarray}\label{leptogen}
- M_i^2P_i^{++}P_i^{--} + \left[ \bar
L_aY_{aa}\tilde \phi\,\ell_{aR} - \mu_{sn}SP_n^{++}P_3^{--} - \mu_n(h^+)^2P_n^{--} -
h_{nab}\bar\ell^c_{aR}\ell_{bR}P_n^{++} + {\rm
H.c.}\right],
\end{eqnarray}
with $i=1,\cdots,3$, $n=1,2$.
%

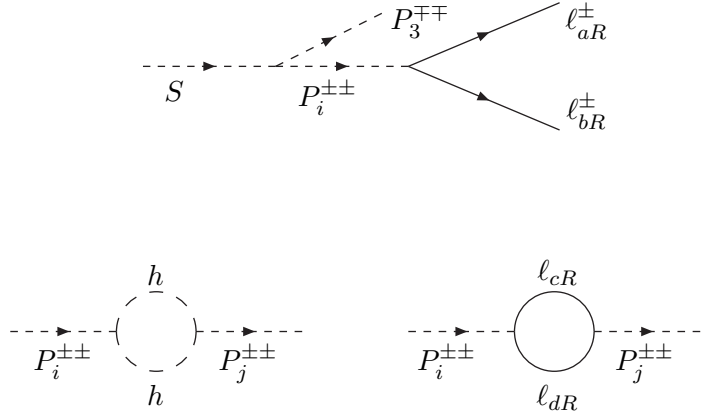
\begin{figure}[ht]
\begin{picture}(300,200)(0,0)
\DashArrowLine(70,150)(120,150){3}\Text(82,145)[t]{$S$}
\DashArrowLine(120,150)(160,170){3}
\Text(175,174)[t]{$P_3^{\mp\mp}$}
\DashArrowLine(120,150)(170,150){3}
\Text(140,145)[t]{$P_i^{\pm\pm}$} \ArrowLine(170,150)(227,174)
\Text(238,174)[t]{$\ell_{aR}^{\pm}$} \ArrowLine(170,150)(227,126)
\Text(238,140)[t]{$\ell_{bR}^{\pm}$}
\DashArrowLine(20,50)(60,50){3} \Text(40,45)[t]{$P_i^{\pm\pm}$}
\DashCArc(75,50)(15,0,180){5} \Text(75,68)[b]{$h$}
\DashCArc(75,50)(15,180,360){5} \Text(75,30)[t]{$h$}
\DashArrowLine(90,50)(130,50){3} \Text(110,45)[t]{$P_j^{\pm\pm}$}
\DashArrowLine(170,50)(210,50){3} \Text(185,45)[t]{$P_i^{\pm\pm}$}
\CArc(225,50)(15,0,180) \Text(227,68)[b]{$\ell_{cR}$}
\CArc(225,50)(15,180,360) \Text(227,30)[t]{$\ell_{dR}$}
\DashArrowLine(240,50)(280,50){3} \Text(260,45)[t]{$P_j^{\pm\pm}$}
\end{picture}
\vspace{-5mm} \caption{
$S\to P_3^{\mp\mp}
\ell_{aR}^{\pm}\ell_{bR}^{\pm}$ at tree level (upper)
and self-energy corrections to the wave functions of $P_i$
(lower) and $i,j=1,2$.}\label{tree}
\end{figure}

As shown in Fig.~\ref{tree},
$CP$ violation occurs in the interference of tree and one-loop
contributions to the decays of
$S\to P_3^{\mp\mp} \ell_{a}^{\pm}\ell_{b}^{\pm}$.
The lepton asymmetry generated in the right-handed
leptons is
transferred to the left-handed ones
since the left-right equilibration for the SM charged leptons
occurs before the sphaleron freezeout~\cite{LRequilibration,Olive1}.
Our result applies so long as
the interactions involving $\ell_R$, which would wash out the
lepton asymmetry before the electroweak phase transition (EWPT), are
strongly suppressed.

For the successful leptogenesis at temperatures $T<M_S$, we require the mass relations $M_3<M_S<M_{i}$ ($i=1,2$) since most of $P_i$ have to be decayed away. Hence their scattering can not wash out the lepton asymmetry. On the other hand, we have introduced a $Z_2$ symmetry to forbid the lepton number violated $\ell\ell P_3$ interactions.

The $S$ decay width at the leading order in $M_S^2/M_{i}^2$ is written as
\begin{eqnarray}
\Gamma_S \simeq \Gamma(S\to P_3^{\mp\mp}\ell^\pm\ell^\pm) = \frac{1}{(2\pi)^3}\frac{1}{96}\frac{1}{4}
\sum_{ijab}\rho_{ab}\mu_{si}\mu_{sj}^*h_{iab}^*h_{jab}\frac{(M_S^2-M_3^2)^2(M_S^2+5M_3^2)}{M_i^2M_j^2M_S^3}
\end{eqnarray}
and at the leading order in $M_3^2/M_S^2$~\cite{Hambye}
\begin{eqnarray}\label{width}
\Gamma_S  = \frac{1}{(2\pi)^3}\frac{1}{96}\frac{1}{4}
\sum_{ijab}\rho_{ab}\mu_{si}\mu_{sj}^*h_{iab}^*h_{jab}\frac{M_S^3}{M_i^2M_j^2},
\end{eqnarray}
where $\rho_{ab}=2-\delta_{ab}$ and $M_S<M_3+2M_h$.
The out-of-equilibrium condition $\Gamma_S<H(T=M_S)$ is satisfied for
\begin{eqnarray}\label{out-of-equilib}
|\mu_{si}h_{jab}| < 640 M_iM_j / \sqrt{\rho_{ab}M_SM_{Planck}},
\end{eqnarray}
where $H =1.66g_*^{1/2}T^2/M_{Planck}$ is the Hubble constant with
$g_*\simeq10^2$ and $M_{Planck} \simeq 10^{19}$~GeV.

The reduced lepton asymmetry
\begin{eqnarray}\label{epsilon}
\epsilon \equiv n_L/n_S = 2[B(S\to\ell\ell P_3^{++}) - B(S\to\ell^c\ell^cP_3^{--})],
\end{eqnarray}
where $n_L= n_l - n_{\bar l}$ with $n_l$, $n_{\bar l}$ and  $n_S$ being the
number densities of leptons, antileptons and $S$,
can be rewritten as~\cite{Hambye}
\begin{eqnarray}\label{asymmetry}
\epsilon &\simeq& 2 A \sum_{ab}\rho_{ab} \left\{ {\rm Im}[\mu_{s1}\mu_{s2}^*h_{1ab}^*h_{2ab}]
\left(\frac{|\mu_1|^2}{M_1^2}-\frac{|\mu_2|^2}{M_2^2}\right)
+ {\rm Im}[\mu_1\mu_2^*\mu_{s1}^*\mu_{s2}]
\left(\frac{|h_{1ab}|^2}{M_1^2}-\frac{|h_{2ab}|^2}{M_2^2}\right)\right. \nonumber \\
&& + {\rm Im}[h_{1ab}h_{2ab}^*\mu_1^*\mu_2]
\left.\left(\frac{|\mu_{s1}|^2}{M_1^2}-\frac{|\mu_{s2}|^2}{M_2^2}\right)\right\}
\end{eqnarray}
with
\begin{eqnarray}\label{A}
A= \frac{1}{\Gamma_S} \frac{1}{(2\pi)^3} \frac{1}{12} \frac{\pi}{(4\pi)^2} \frac{1}{4} \frac{M_S^3}{M_1^2M_2^2}
=\frac{1}{2\pi M_1^2M_2^2}\left( \sum_{ijab}\frac{\rho_{ab}\mu_{si}\mu_{sj}^*h_{iab}^*h_{jab}}{M_i^2M_j^2} \right)^{-1}.
\end{eqnarray}

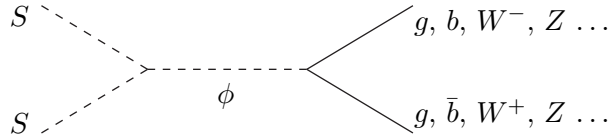
\begin{figure}[ht]
\begin{picture}(300,100)(0,0)
\DashLine(60,74)(100,50){3} \Text(52,74)[t]{$S$}
\DashLine(60,26)(100,50){3} \Text(52,26)[b]{$S$}
\DashLine(100,50)(160,50){3} \Text(130,45)[t]{$\phi$}
\Line(160,50)(200,74) \Line(160,50)(200,26)
\Text(239,74)[t]{$g$, $b$, $W^-$, $Z$ \dots}
\Text(239,28)[b]{$g$, $\bar b$, $W^+$, $Z$ \dots}
\end{picture}
\vspace{-5mm} \caption{
Scattering $SS\leftrightarrow$~all.}\label{scattering}
\end{figure}

The time evolution of $\epsilon$ depends on $M_S$ and the temperatures $T_*^a\ (a=e,\mu,\tau)$ of the left-right
equilibrations for the corresponding charged leptons. These temperatures are
determined by the equilibrium condition $\Gamma_{\phi}\geq
H$~\cite{Olive1}, where $\Gamma_{\phi}$ denotes the width of the SM Higgs boson decay of
$\phi\to \bar L_a \ell_{aR}$.
One finds that $T_*^e\sim 10^4~{\rm GeV}\ll T_*^\mu \ll T_*^\tau$.
The Boltzmann equation for the lepton asymmetry \cite{Ma_Sarkar,evolution,KT,Luty,DZ} is given by
\begin{eqnarray}\label{BE1}
    \frac{dn_L}{dt} + 3Hn_L = \frac{\epsilon}{2}\langle\Gamma_S\rangle (n_S-n_S^{eq})
    - \langle\Gamma_S\rangle\left(\frac{n_S^{eq}}{n_\gamma}\right)n_L -
    2\langle\sigma| {\bm v}|\rangle n_\gamma n_L,
\end{eqnarray}
which is the same as that with the initial left-handed lepton
asymmetry, where $n_S^{eq}$ is the equilibrium
number density of $S$, $n_\gamma$ is the photon density,
$\langle\ \rangle$ represents thermal averaging,
${\bm v}$ is
the relative velosity of the incoming particles, and
$\sigma=\sigma(\ell^{\mp}\ell^\pm P_3^{\pm\pm}\to \ell^\pm\ell^\pm P_3^{\mp\mp})$. Here, we have assumed
$T<2M_h$ and the $CPT$ invariance.
The density of $S$ satisfies
\begin{eqnarray}\label{BE2}
    \frac{dn_S}{dt} + 3Hn_S = -\langle\Gamma_S\rangle(n_S-n_S^{eq}) -\langle\sigma_{s}|{\bm v}|\rangle(n_S^2-n_S^{eq2}),
\end{eqnarray}
where $\sigma_{s}$ is the cross section of the scattering processes $SS\to\phi\to$~all, shown in Fig.~\ref{scattering}.
%
In terms of the dimensionless variable $x\equiv M_S/T$, the
particle densities per entropy density of $L\equiv n_L/s$ and
$Y\equiv n_1/s$ with $s\simeq7n_\gamma$ in the present epoch, using the relation $t=x^2/(2H(x=1))$, the
Boltzmann equations in Eqs. (\ref{BE1}) and  (\ref{BE2}) can be
rewritten as
\begin{eqnarray}\label{BE3}
    &&\frac{dL}{dx} = \frac{\epsilon}{2}K(Y-Y^{eq})x - \gamma KLx,\\
    &&\frac{dY}{dx} = -K(Y-Y^{eq})x -\frac{\gamma_{s}}{sH(x=1)}(Y^2-Y^{eq2}),\label{BE4}
\end{eqnarray}
respectively, where $K\equiv\langle\Gamma_S\rangle/H(x=1)$ is the
effectiveness of the decays in the crucial epoch ($T\sim M_S$),
$\gamma \equiv g_*Y^{eq} + 2\langle\sigma|{\bm v}|\rangle n_\gamma/\langle\Gamma_S\rangle$ and the reaction density for the scattering processes~\cite{Hambye,HMS}
\begin{eqnarray}
    \gamma_s = \frac{T}{64\pi^2} \int\limits_{4M_S^2}^\infty ds \hat\sigma_s(s)\sqrt{s}K_1(\sqrt{s}/T)
\end{eqnarray}
with the reduced cross section $\hat\sigma_s$ given by $2(s-4M_S^2)\sigma_S(s)$ and the  modified Bessel function of $K_1$. Since the dependence of the lepton asymmetry on the magnitude of the scattering is much slower than the linear one, we make the following estimate
\begin{eqnarray}
    \sigma_s = \frac{1}{\pi\sqrt{s}}\frac{1}{\sqrt{s-4M_S^2}}\left( \frac{\lambda_{\phi s}v}{M_Z}\right)^2
\end{eqnarray}
with the SM Higgs vev $v=246$~GeV. The scattering term is negligible for small (large) values of $\lambda_{\phi s}$ ($M_S$). By requiring that the effect of this term on the evolution of the lepton asymmetry is small we have the bounds
\begin{eqnarray}
    &\lambda_{\phi s}< 10^{-5}\quad &{\rm for}\quad M_S\sim 1~{\rm TeV},\nonumber\\
    &\lambda_{\phi s}< 10^{-4}\quad &{\rm for}\quad M_S\sim 10^2~{\rm TeV}.
\end{eqnarray}

In the range of $K\ll 1$, one obtains $L = \epsilon/(2g_*)$~\cite{KT}.
In the EWPT, the lepton asymmetry
in our model is converted to the net baryon asymmetry per
entropy density
\begin{eqnarray}
    B\equiv \frac{n_B}{s}\equiv \frac{n_b-n_{\bar b}}{s}
\end{eqnarray}
due to
the relation \cite{Harvey_Turner,BPY,H-Sh}
\begin{eqnarray}\label{n_B}
    B_f = \frac{28}{79} (B-L),
\end{eqnarray}
where the index $f$ represents the present value and there is no  initial baryon asymmetry.
Note that we have ignored
the temperature dependence of the $n_B$ vs. $n_L$ for
$T\sim v$~\cite{BPY,Laine}.
From Eq.~(\ref{asymmetry}), without loss of generality
we obtain
\begin{eqnarray}\label{BA1}
    n_B/n_\gamma = 10^{-2}\epsilon\,.
\end{eqnarray}

Since we have the additional DCH the parameters $h_{iab}\equiv h_0$~\cite{Nebot} and $\mu_i\equiv\mu$ in our model are more relaxed
than in the ZB model~\cite{Nebot}, given by
\begin{eqnarray}\label{normal}
    \frac{0.42}{\sqrt{\kappa}}~{\rm TeV} \leq M_\alpha < 10^3 \kappa^4~{\rm
    TeV}, \quad \frac{0.1}{\kappa}~{\rm TeV} < \mu < 10^3 \kappa^5~{\rm
    TeV}, \quad \frac{0.01}{\kappa^2} \leq h_0 \leq \kappa;
\end{eqnarray}
and
\begin{eqnarray}\label{inverted}
    \frac{0.78}{\sqrt{\kappa}}~{\rm TeV} \leq M_\alpha <
    274 \kappa^4~{\rm TeV}, \quad \frac{0.36}{\kappa}~{\rm TeV} < \mu < 274 \kappa^5~{\rm
    TeV}, \quad \frac{0.036}{\kappa^2} \leq h_0 \leq
    \kappa;
\end{eqnarray}
for the normal and inverted hierarchies of the neutrino masses,
respectively, where the parameter $\kappa$ $(\geq \mu/M_h)$ is
taking to be $\sim 1$ and $\alpha=h^\pm$ and
$P_i^{\pm\pm}$. Taking the central values $h_0\sim 0.1$ and $\mu\sim 1$~TeV and the neutral singlet mass $M_S\sim 1$~TeV,
we get $\epsilon\sim 1\, {\rm TeV}^2M_1^{-2}$ ($M_1 < M_2$) and
\begin{eqnarray}\label{out-of-equilib1}
    \mu_s\equiv\mu_{si}\lesssim 10^{-4}~{\rm TeV}^{-1} M_1^2
\end{eqnarray}
in Eq.~(\ref{out-of-equilib}) with $\mu_s$ and $M_1$ in TeV.
It is easy to satisfy the condition in Eq.~(\ref{out-of-equilib1})
and describe observed baryon number asymmetry $n_B/n_\gamma = 6\times 10^{-10}$
~\cite{WMAP} since $\mu_s$ is free parameter of the model and for the DCH mass we only require $M_1>M_S$.

We note that one may consider a minimal extension of this model by including the second neutral singlet scalar $S_0$ to generate the dark matter in the universe. $S_0$ particles are stable if their masses satisfy the condition
$$M_{S_0}<M_{P_3}+2m_e.$$
Moreover, the relation $M_S<3M_{S_0}$ is needed to prevent the $S\to 3S_0$ decays which can deplete the lepton asymmetry. If both these conditions are satisfied, $S_0$ can be the ordinary candidate for stable dark matter particles~\cite{Pospelov}.

In conclusion, we have investigated a new mechanism for the
leptogenesis in the extension of the ZB  model. We have shown
that the observed baryon number asymmetry of the universe can be
produced through the decay of the neutral scalar $S$ for both normal and
inverted hierarchies of the neutrino masses. This $S$ boson at the TeV scale can
be probed directly at the near future colliders. One of the advantages
of the our mechanism is that the degeneracy for
the scalar masses and the hierarchy of couplings is not  required.
We have also pointed out that the dark matter can be generated if we include a second neutral scalar in our model.

\section*{Acknowledgements} This work is financially supported by
the National Science Council of Republic of China under the
contract \#: NSC-95-2112-M-007-059-MY3. We would like to thank Pei-Hong~Gu and A.~V.~Borisov
for helpful discussions.


\end{document}